# Projecting the Forward Rate Flow onto a Finite Dimensional Manifold

Erhan Bayraktar [*]    Li Chen [†]    H. Vincent Poor [‡]

**Abstract**

Given a Heath-Jarrow-Morton (HJM) interest rate model $\mathcal{M}$ and a parametrized family of finite dimensional forward rate curves $\mathcal{G}$, this paper provides a technique for projecting the infinite dimensional forward rate curve $r_t$ given by $\mathcal{M}$ onto the finite dimensional manifold $\mathcal{G}$. The Stratonovich dynamics of the projected finite dimensional forward curve are derived and it is shown that, under the regularity conditions, the given Stratonovich differential equation has a unique strong solution. Moreover, this projection leads to an efficient algorithm for implicit parametric estimation of the infinite dimensional HJM model. The feasibility of this method is demonstrated by applying the generalized method of moments.

## 1 Introduction

Consider a complete filtered probability space $(\Omega, \mathcal{F}, (\mathcal{F}_t), \mathbb{P})$ satisfying the usual conditions, and where $\mathbb{P}$ denotes the risk-neutral measure. Consider an HJM model $\mathcal{M}$ (see [9]) with the Musiela parametrization (see [11]). The dynamics of the forward rate curve under $\mathbb{P}$ can be given by the following infinite dimensional stochastic differential equation (SDE):

$$dr_t(x) = \tilde{\mu}(r_t, x)dt + \sigma(r_t, x)dW_t, \qquad (1.1)$$

where $W$ is an $m$-dimensional standard $\mathbb{P}$-Brownian motion. From the arbitrage-free condition given by [9], it follows that

$$\tilde{\mu}(r_t, x) = \frac{\partial}{\partial x} r_t(x) + \sigma(r_t, x) \int_0^x \sigma(r_t, u)^T du.$$

[*] Department of Mathematics, University of Michigan, 2074 East Hall, Ann Arbor, MI 48109-1109, email: erhan@umich.edu

[†] Lehman Brothers, Fixed Income Derivatives Research, 745 Seventh Avenue, New York, NY 10019 email: lichen@lehman.com

[‡] Department of Electrical Engineering, Princeton University, Princeton, NJ 08544, email: poor@princeton.edu

Since (1.1) is an infinite dimensional stochastic differential equation, it is difficult in practice to apply it to characterize the dynamics of the forward rate. More commonly, simpler parametric families of curves are used to characterize the forward rate curve. This practice gives rise to the following issues:

- Under what conditions, if any, does the infinite dimensional forward rate curve $r_t(\cdot)$ defined in (1.1) admit a finite-dimensional realization?

- Under what conditions is this forward curve consistent with a given family of finite-dimensional curves?

Under the assumption of smoothness of the volatility function $\sigma$, Björk *et al.* ([2],[3]) give necessary and sufficient conditions for the consistency problem, and for the existence of a finite-dimensional realization. Their results are developed further by Filipović and Teichmann ([7]). (See also Filipović [6] and Bayraktar et. al. [1].) In this paper however, we will assume that the forward rate curve does not admit a finite dimensional realization. It is of interest to find the projection of the infinite dimensional forward rate curve onto a finite dimensional manifold, and we consider this problem here. Precisely, starting from the specified finite dimensional manifold, each time the velocity vector points out of the this manifold, it is projected onto the tangent space. The velocity vector field obtained by this procedure now describes a flow which is finite dimensional.

When the finite dimensional manifold is linear this projection characterizes the behavior of constructing forward rate curves from cross sectional market data by interpolation using a parameterized family of finite dimensional curves. Using our results this data can be used to calibrate an infinite dimensional HJM model (see also Remark 3.2) by deriving the coordinate dynamics of the projected forward curve provides an efficient way for implicitly estimating the evolution of the original HJM model.

The remainder of the paper is organized as follows. In Section 2, we introduce a fundamental mathematical formalism within which we consider this problem. In particular, we confine our discussion to the case in which the solution $r_t(\cdot)$ given by (1.1) lies in a specific Sobolev space with a finite time horizon. Then in Section 3, we derive the dynamics of the coordinates of the projected forward curve given a finite dimensional manifold $\mathcal{G}$. In Section 4, we propose a simple and efficient strategy for parametric estimation of the original HJM model through the projected forward curves. In particular, we demonstrate that the generalized method of moments can be straightforwardly implemented in this situation.

## 2 Basic Mathematical Formalism and Projection

To ensure sufficient regularity of model (1.1), we restrict attention to the situation in which the solution $r_t(\cdot)$ lie within a Sobolev space $\mathcal{B}$ with a finite time horizon $T_0$ defined as the space of all infinitely differentiable functions

$$r : [0, T_0] \mapsto \mathbb{R}$$

having finite norm

$$||r||_\mathcal{B} := \sum_{n=0}^{\infty} 2^{-n} \int_0^{T_0} \left(\frac{d^n r}{dx^n}(x)\right)^2 dx < \infty.$$

**Remark 2.1** *In order to include all constant curves into the space, there usually exists an exponential weighting function $e^{-\gamma x}(\gamma > 0)$ in the definition of the norm for a weighted Sobolev space. However, in our case, since we already confine us in a finite time horizon, this weighting function becomes unnecessary.*

Now let us consider the forward rate dynamics $r_t(\cdot)$ satisfying Equation (1.1). By using the infinite dimensional Itô formula (see [4]), (1.1) can be written in Stratonovich form as

$$dr_t(x) = \left(\tilde{\mu}(r_t, x) - \frac{1}{2}[\sigma'_r(r_t)(\sigma(r_t))](x)\right) dt + \sigma(r_t, x) \circ dW_t, \quad (2.1)$$

where $\circ$ denotes the Stratonovich integral and

$$\sigma'_r(r_t)\sigma(r_t) := \sum_{i=1}^{m} \sigma'_{ir}(r_t)\sigma_i(r_t).$$

Here $\sigma'_r$ denotes the Fréchet derivative of the vector field $\sigma$ with respect to $r_t$. For convenience, let us denote

$$\mu(r_t) := \tilde{\mu}(r_t) - \frac{1}{2}\sigma'_r(r_t)\sigma(r_t). \quad (2.2)$$

The following assumption guarantees the existence a local strong solution for (1.1) in the space $\mathcal{B}$.

**Assumption 2.1** *i) $\sigma$ is a smooth mapping;*
  *ii) the mapping $r \mapsto \sigma(r) \int_0^\cdot \sigma(r, u)^T du - \frac{1}{2}\sigma'_r(r)(\sigma(r))$ is also smooth.*

By a smooth mapping, we mean this mapping belongs to $C^\infty$.

**Remark 2.2** From Assumption 2.1, it is straightforward to deduce that

$$\mu(r,\cdot) \in \mathcal{B}, \quad \sigma(r,\cdot) \in \mathcal{B}, \quad \forall\, r \in \mathcal{B}.$$

Therefore, within a regular space $\mathcal{B}$, we can start to discuss the projection of the forward rate flow $r_t$ onto a certain finite dimensional manifold.

Let $\mathcal{G} := \{G(\vec{z},\cdot) \in \mathcal{B} : \vec{z} = (z^1,...,z^n) \in \mathbb{Z} \subset \mathbb{R}^n\}$ be a finite dimensional manifold, where $\mathbb{Z}$ is the *parameter space*. Consider a curve $\beta$ in this manifold of the form $\beta : h \mapsto G(\vec{z}(h),\cdot)$, where $h \mapsto \vec{z}(h)$ is an $n$-dimensional curve. Then, by the chain rule, we can compute the following Fréchet derivative:

$$D\beta(0) = DG(\vec{z}(h),\cdot)|_{h=0} = \sum_{k=1}^{n} \frac{\partial G(\vec{z},\cdot)}{\partial z^k} \dot{z}^k(0),$$

Therefore, for each $\vec{z}_0$ in $\mathbb{Z}$, the tangent space in this manifold is then obtained by considering all the curves passing through $G(\vec{z}_0,\cdot)$ and tangential to $\mathcal{G}$. From this argument it is clear that the tangent vector space is given by

$$\mathcal{S}_{\vec{z}_0} := span\left\{ \frac{\partial G(\vec{z},\cdot)}{\partial z^1}, ..., \frac{\partial G(\vec{z},\cdot)}{\partial z^n} \right\}\bigg|_{\vec{z}=\vec{z}_0}.$$

Without loss of generality, it is assumed that $\frac{\partial G(\vec{z},\cdot)}{\partial z^1}, ..., \frac{\partial G(\vec{z},\cdot)}{\partial z^n}$ are linearly independent.

Let us now consider orthogonal projections between any linear space $V \in \mathcal{B}$ and the tangent vector space $\mathcal{S}_{\vec{z}_0}$. Since the generating basis of $\mathcal{S}_{\vec{z}_0}$ is not necessarily orthogonal, we have the following projection formula $\Pi_{\vec{z}} : V \mapsto span\left\{ \frac{\partial G(\vec{z},\cdot)}{\partial z^1}, ..., \frac{\partial G(\vec{z},\cdot)}{\partial z^n} \right\}$,

$$v \to \sum_{i=1}^{n} \left[ \sum_{j=1}^{n} \bar{\lambda}_{ij} \left\langle v, \frac{\partial G(\vec{z},\cdot)}{\partial z^j} \right\rangle \right] \frac{\partial G(\vec{z},\cdot)}{\partial z^i}, \qquad (2.3)$$

where $(\bar{\lambda}_{ij}) := \Lambda^{-1}$ with $\Lambda := \left( \langle \frac{\partial G(\vec{z},\cdot)}{\partial z^i}, \frac{\partial G(\vec{z},\cdot)}{\partial z^j} \rangle \right)$. Here $\langle \cdot, \cdot \rangle$ denotes the inner product of the projection. It is worth mentioning that this inner product is not necessarily the one defined in the space $\mathcal{B}$, which was constructed so as to guarantee the boundedness of $\partial/\partial x$, together with Assumption (2.1)(ii) assures the smoothness of $\tilde{\mu}$. Actually an appropriate choice of this inner product should reflect the calibration criteria of a financial institution. Generally we can define a Hilbert space $\mathcal{H}$ with a finite time horizon $T_0$ containing all the functions

$$v : [0, T_0] \mapsto \mathbb{R}$$

satisfying $||v||_{\mathcal{H}} < \infty$ with norm $||\cdot||_{\mathcal{H}}$ defined as

$$||v||_{\mathcal{H}} := \int_0^{T_0} v^2(x)w(x)dx, \quad (2.4)$$

where $w(\cdot)$ is a positive bounded continuous function on $[0, T_0]$. The corresponding inner product is defined by

$$\langle u, v \rangle = \int_0^{T_0} u(x)v(x)w(x)dx, \quad \forall\ u, v \in \mathcal{H}. \quad (2.5)$$

Typically, fund managers, when interpolating the forward curve by using a certain curve family would require more accuracy in the long term region, and accordingly would choose choose $w(x)$ to give higher weight to larger values of $x$. Alternatively, traders would choose $w(x)$ to put greater weight on the short term.

**Remark 2.3** *It follows from (2.4) that $\mathcal{B} \subset \mathcal{H}$.*

**Assumption 2.2** *We assume that $G(\vec{z}, \cdot) \in \mathcal{B}$.*

**Remark 2.4** *By Remark 2.2 and Assumption 2.2, we have that*

$$\sigma(G(\vec{z}_t, \cdot)) \in \mathcal{B}, \quad \mu(G(\vec{z}_t), \cdot) \in \mathcal{B}$$
$$and \quad \frac{\partial G(\vec{z}_t, \cdot)}{\partial z^i} \in \mathcal{H}, \quad \forall\ i = 1, 2, ..., n.$$

*Therefore, we can discuss the projection without worrying about its existence.*

The projected forward rate flow of $r_t$ onto a finite manifold $\mathcal{G}$ is thus given by
$$dG(\vec{z}_t, \cdot) = \Pi_{\vec{z}_t}\left[\mu(G(\vec{z}_t, \cdot))\right] dt + \Pi_{\vec{z}_t}\left[\sigma(G(\vec{z}_t, \cdot))\right] \circ dW_t. \quad (2.6)$$
where $\Pi_{\vec{z}}$ is defined by (2.3).

# 3 The Dynamics of Finite Dimensional Coordinates

In order to derive the diffusion process for $\vec{z}$, first we give the following lemma.

**Lemma 3.1** *For any two continuous semimartingales $X_t$ and $Y_t$ having finite quadratic variation on finite intervals, i.e. $[X]_T < \infty$, and $[Y]_T < \infty$ for each*

$T > 0$. Then for $\forall\, T > 0$, we have

$$\int_0^T (X_t Y_t) \circ dW_t = \int_0^T X_t \circ (Y_t \circ dW_t), \quad a.s.. \tag{3.1}$$

*Proof.* From the definitions of Stratonovich's integral and variation, it is sufficient to prove that

$$\lim_{||\Psi|| \to 0} \sum_{i=1}^n (X_{t_i} - X_{t_{i-1}})(Y_{t_i} - Y_{t_{i-1}})(W_{t_i} - W_{t_{i-1}}) = 0, \quad a.s., \tag{3.2}$$

where $\Psi = \{t_0, t_1, ..., t_n\}$ is a partition of $[0, T]$ and $||\Psi||$ is the mesh of this partition, namely, $||\Lambda|| = \max_{1 \leq i \leq n}\{|t_i - t_{i-1}|\}$. Since for any $\Lambda$, we have

$$\left| \sum_{i=1}^n (X_{t_i} - X_{t_{i-1}})(Y_{t_i} - Y_{t_{i-1}})(W_{t_i} - W_{t_{i-1}}) \right| \leq \max_{1 \leq i \leq n} |W_{t_i} - W_{t_{i-1}}| \times$$

$$\sqrt{\sum_{i=1}^n (X_{t_i} - X_{t_{i-1}})^2} \sqrt{\sum_{i=1}^n (Y_{t_i} - Y_{t_{i-1}})^2}, \tag{3.3}$$

by the uniform continuity of the Brownian motion $W$ on the compact support $[0, T]$ and the finiteness of $[X]_T$, and $[Y]_T$, it follows that (3.2) is true. Lemma 3.1 follows. $\square$

By applying the Stratonovich chain rule, we can derive that

$$dG(\vec{z}_t, \cdot) = \sum_{i=1}^n \frac{\partial G(\vec{z}_t, \cdot)}{\partial z^i} \circ dz_t^i. \tag{3.4}$$

On the other hand, by (2.3) and (2.6), we also have

$$dG(\vec{z}_t, \cdot) = \sum_{i=1}^n \left[ \sum_{j=1}^n \bar{\lambda}_{ij} \left\langle \mu(G(\vec{z}_t, \cdot)), \frac{\partial G(\vec{z}_t, \cdot)}{\partial z^j} \right\rangle \right] \frac{\partial G(\vec{z}_t, \cdot)}{\partial z^i} dt$$

$$+ \sum_{i=1}^n \left[ \sum_{j=1}^n \bar{\lambda}_{ij} \left\langle \sigma(G(\vec{z}_t, \cdot)), \frac{\partial G(\vec{z}_t, \cdot)}{\partial z^j} \right\rangle \right] \frac{\partial G(\vec{z}_t, \cdot)}{\partial z^i} \circ dW_t \tag{3.5}$$

By comparing (3.4) and (3.5) and using Lemma 3.1, we conclude that the finite dimensional vector $\vec{z}$ follows a diffusion process:

$$\begin{aligned}
dz_t^i &= \left[\sum_{j=1}^n \bar{\lambda}_{ij} \left\langle \mu(G(\vec{z}_t, \cdot)), \frac{\partial G(\vec{z}_t, \cdot)}{\partial z^j} \right\rangle \right] dt \\
&+ \left[\sum_{j=1}^n \bar{\lambda}_{ij} \left\langle \sigma(G(\vec{z}_t, \cdot)), \frac{\partial G(\vec{z}_t, \cdot)}{\partial z^j} \right\rangle \right] \circ dW_t, \quad \forall\, i \in \{1, ..., n\}.(3.6)
\end{aligned}$$

**Theorem 3.1** *Suppose $G$ satisfies Assumption 2.2, and that Assumption 2.1 is valid. Then there exists a unique strong solution for the SDE (3.6).*

*Proof.* This result follows directly from Remark 2.2 and Proposition 5.2.21 in [10].

**Remark 3.1** *Each trading day banks perform a yield curve fitting to determine the prices of bonds with maturities whose prices are not observed because they are not traded on the market. This is done because the payoffs of even vanilla products such as caps, floors and swaptions consists of a sequence of cash flows. And the valuation of several products like these require the entire yield curve. Each trading day the prices of several bonds are set by the yield curve obtained by curve fitting. Assume now that the prices between the two calibrations follow an HJM model (this models the market activity between the times of curve fitting). We will show the calibration of the banks can be obtained by discretizing (2.6), i.e., we will show that $G(\vec{z}_{\Delta t}, \cdot)$ given by (2.6) in fact minimizes*

$$\min_{G(\vec{y}_{\Delta t}, \cdot) \in \mathcal{G}} \|r_{\Delta t} - G(y\vec{\Delta}t, x)\|_{\mathcal{H}}. \tag{3.7}$$

*Therefore using the results of next section one can calibrate an infinite dimensional HJM model using the yield curve data obtained form weigted linear regression.*

The minimization in (3.7) is equivalent to showing

$$\langle r(\Delta t) - G(z_{\Delta t} \cdot), G_z(z_{\Delta t}, \cdot) \rangle_{\mathcal{H}} = 0, \tag{3.8}$$

where $G_z$ is the derivative of $G$ with respect to $\vec{z}$. Assume that $r_0 \in \mathcal{G}$ and take let the SDE in (2.6) also start from this point. Then discretizing (2.1) we obtain

$$r_{\Delta t} = r_0 + \mu(r_0, x)\Delta t + \sigma(r_0, x)W_{\Delta t}.$$

*We can apply the same discretization to (2.6) and obtain*

$$r_{\Delta t} - G(\vec{z}_{\Delta t}, \cdot) = (\mu(r_0, x) - \Pi_{\vec{z}_0}\left[\mu(G(\vec{z}_0, \cdot))\right])\Delta t + (\sigma(x) - \Pi_{\vec{z}_0}\left[\sigma(x)\right])W_{\Delta t}. \tag{3.9}$$

*By the definition of the projection operator $\Pi_{\vec{z}_0}$ the right-hand-side is perpendicular to the $\mathcal{S}_{\vec{z}_0}$ therefore so is the left-hand-side, hence (3.8) is satisfied.*

**Remark 3.2** *Consider the following affine term structure:*

$$G(\vec{z}_t, x) = g_0(x) + g_1(x)z_t^1 + ... + g_n(x)z_t^n \tag{3.10}$$

*Then the minimizer in (3.7) is guaranteed to be unique, since $\mathcal{G}$ is linear.*

# 4 Parametric Estimation of the Original HJM Models

In this section, we discuss parametric estimation of the original HJM model $\mathcal{M}$ by using its projection onto a finite dimension manifold $\mathcal{G}$. Generally speaking, estimating parameters of an infinite dimensional SDE is a difficult task to implement. However, since we have already derived the SDE (3.6) governing the dynamics of its projected finite dimensional vector $\vec{z}$, this provides us an easier way to empirically investigate the underlying HJM model.

Suppose $\theta$ is a vector of parameters parametrizing the HJM model. By projecting this forward curve into a finite dimensional manifold $\mathcal{G}$, we have the following general form for the diffusion process of $\vec{z}$:

$$d\vec{z}_t = A(\vec{z}_t, \theta) + B(\vec{z}_t, \theta) \circ dW_t, \tag{4.1}$$

where $W$ is an $m$-dimensional standard Brownian motion. It will be convenient to write (4.1) in Itô form which is as follows.

$$d\vec{z}_t = \left(A(\vec{z}, \theta) + \frac{1}{2}\sum_{j=1}^m \left(\frac{\partial B^j}{\partial \vec{z}}(\vec{z}_t, \theta)\right)B^j(\vec{z}_t, \theta)\right)dt + B(\vec{z}_t, \theta)dW_t, \tag{4.2}$$

where $B^j$ denotes the $j$th column of $B$.

Here we apply the generalized method of moments (GMM) proposed by Hansen (1982 [8]) to estimate $\theta$. For banks, periodic calibration of the initial forward curve produces a time series $\{\vec{z}_{t_k}\}_{1 \le k \le N}$. Let $\Delta := t_{k+1} - t_k$, for each

$k = 1, ..., N$. By discretizing (4.1), we obtain a discrete-time model:

$$\vec{z}_{t_{k+1}} - \vec{z}_{t_k} = \left( A(\vec{z}_{t_k}, \theta) + \frac{1}{2} \sum_{j=1}^{m} \left( \frac{\partial B^j}{\partial \vec{z}}(\vec{z}_{t_k}, \theta) \right) B^j(\vec{z}_{t_k}, \theta) \right) \Delta + B(\vec{z}_{t_k}, \theta) \epsilon_k, \tag{4.3}$$

for $k = 1, 2, ..., N$, where $\epsilon_k$ is a $m$-dimensional Gaussian random vector with mean 0 and covariance matrix $I\Delta$, i.e., $\{\epsilon_k\}_{1 \leq k \leq N}$ are mutually independent.

Since $z_{t_k}$ and $\epsilon_k$ are independent we construct the moment functions $\{h_k(\theta)\}$ as

$$h_k(\theta) = \vec{z}_{t_{k+1}} - \vec{z}_{t_k} - \left( A(\vec{z}_{t_k}, \theta) + \frac{1}{2} \sum_{j=1}^{m} \left( \frac{\partial B^j}{\partial \vec{z}}(\vec{z}_{t_k}, \theta) \right) B^j(\vec{z}_{t_k}, \theta) \right) \Delta, \tag{4.4}$$

for $k = 1, ..., N$ and denote the sample average by $f_N(\theta) := \frac{1}{N} \sum_{k=1}^{N} h_k(\theta)$. Then by

$$\hat{\theta}_N = \min_\theta \{\langle f_N(\theta), f_N(\theta) \rangle\}, \tag{4.5}$$

is the least squares estimator of $\theta$. Under fairly general conditions (see [8]), the estimator $\hat{\theta}_N$ offers a consistent estimator of $\theta_0$.

**Remark 4.1** *If the dimension of $\theta$ is high, it is straightforward to strengthen this algorithm by adding more moment functions.*

As argued by Hansen in [8], the estimator (4.5) is generally not efficient as far as its convergence rate is concerned. The least squares estimator of (4.5) can be improved by taking a weighted least squares estimator. Suppose that $\{h_k(\theta_0)\}$ is strictly stationary, where $\theta_0$ is the true parameter value, and define

$$\Gamma_\nu := E\{h_k(\theta_0) h_{k-\nu}(\theta_0)'\}. \tag{4.6}$$

Assuming $\{\Gamma_\nu\}$ is absolutely summable we define

$$S := \sum_{\nu=-\infty}^{\infty} \Gamma_\nu. \tag{4.7}$$

The optimal GMM estimator is given by

$$\hat{\theta}_N^* = \min_\theta \{f_N(\theta) S^{-1} f_N^T(\theta)\}. \tag{4.8}$$

To implement (4.8) we need an inital estimate of $S$, which calls for an initial estimate $\theta$. The least-squares estimate (4.5) gives such an estimate, which can be used to calculate an estimate for $S$ as follows([12]):

$$S_N = \hat{\Gamma}_{0,N} + \sum_{\nu=1}^{q} \left(1 - (\nu/(q+1))\right)\left(\hat{\Gamma}_{\nu,N} + \hat{\Gamma}_{\nu,N}\right) \quad (4.9)$$

where

$$\hat{\Gamma}_{\nu,N} = \frac{1}{N} \sum_{n=\nu+1}^{N} h_n(\hat{\theta}) h_{n-\nu}(\hat{\theta})' \quad (4.10)$$

and where $\hat{\theta}$ is the estimator obtained using (4.5). This estimate of $S$ is then used to compute an estimate of $\theta$ using (4.8). This obvious recursion can then be carried until the estimator becomes stable.

## Acknowledgements

The work was supported in part by the Office of Naval Research under Grant N00014-03-1-0102.